\documentclass[aps,prl,preprint,onecolumn,amsmath,amssymb,groupedaddress]{revtex4-2}
\usepackage{graphicx}
\usepackage{chemformula}
\usepackage{bm}
\usepackage{physics}
\usepackage{upgreek}
\usepackage{titlesec}
\usepackage{xfrac}
\usepackage{amsmath}
\usepackage{hyperref}

\titleformat{\section}[hang]{\normalfont\large\bfseries}{\thesection}{1em}{}
\titleformat{\subsection}[hang]{\normalfont\normalsize\bfseries}{\thesubsection}{1em}{}
\titleformat{\subsubsection}[hang]{\normalfont\normalsize\itshape}{\thesubsubsection}{1em}{}

\begin{document}

\title{
Electrically-driven chiral emission from plasmonic tunnel junctions 
}

\author{Yuanyang Xie*$^\#$}
\author{Alexey V. Krasavin$^\#$}
\author{Anatoly V. Zayats*}

\affiliation{Department of Physics and London Centre for Nanotechnology, King's College London, London, WS2R 2LS, UK \\
*Email Address: yuanyang.xie@kcl.ac.uk, a.zayats@kcl.ac.uk,$^\#$These authors contributed equally to this work.}

\date{\today}
\clearpage
\begin{abstract}

Chirality plays a crucial role in a broad range of processes including light-matter interactions in physics, chemistry and biology, which opens up new applications in nanophotonics, quantum technologies and photochemistry. Quantum tunnelling 
provides a promising mechanism for light generation at the nanoscale, however the realisation of chiral light emission has remained elusive. Here, by integrating  tunnel junctions with chiral plasmonic nanohelicoids, we achieve nanoscale generation of chiral light at a single-particle level. 
The tunnelling-driven resonant excitation of chiral dipolar modes of the nanohelicoids results in emission of a vortex light beam possessing both spin angular momentum with handedness selectivity of over 0.8 and its orbital counterpart, equal in magnitude and opposite in sign.
The developed approach offers a new means for sculpturing photon spin generation at the nanoscale, highlighting its potential for next-generation optical components in display and AR/VR applications, as well as quantum information processing and photochemistry.

\end{abstract}
 
\maketitle
\clearpage

Chirality is a fundamental symmetry property where an object cannot be superimposed on its mirror image, a distinction with profound consequences across chemistry, biology, and physics. This fundamental property of matter finds a direct counterpart in electromagnetic waves: a chiral light beam carries intrinsic angular momentum, including a spin angular momentum (SAM) defining its handedness and an orbital angular momentum (OAM) defining the phase variation around the beam axis. Such chiral states of light enable selective interaction with chiral matter and drive advanced applications from display technologies and sensing to quantum information processing and topological photonics \cite{wan2023anomalous,chen2023compact,furlan2024chiral,furlan2025electrical,aita-conversion}. Generation of chiral light at the nanoscale is currently a major focus for applications and is approached in both optical and electric excitation regimes \cite{nguyen2023large}.

To induce chirality in light waves, magnetised materials, chiral molecules or chiral structures have been exploited \cite{nishizawa2017pure,zinna2017design,zhang2022chiral,jiang2023circular}. However, the requirements of an external magnetic field, complex designs, and a weak chiral response of materials impede the miniaturisation of components for chiral light emission \cite{dorrah2022tunable,nishizawa2017pure,furlan2024chiral}. Chiral plasmonic nanostructures offer unique advantages featuring tuneable field enhancement, particularly related to an increased local density of chiral optical states (chiral LDOS) which imprints chirality in the emission from achiral emitters \cite{li2025chiral,wang2019induced, liang2025room,zheng2025circularly}. Usually the emitters are presented by molecules or quantum dots which are excited optically. At the same time, electrically-driven nanoscale chiral light generation would be much more advantageous, particularly in practical applications.

The light emission from tunnel junctions represents a distinct form of electroluminescence with a controllable and broadband spectrum \cite{wang2024upconversion}. Unlike traditional mechanisms based on semiconductor p-n junctions, this electrically-driven process relies on inelastic electron tunnelling within a metal-insulator-metal (MIM) heterostructure \cite{parzefall2018light}. In the process of quantum tunnelling, some fraction of the electrons tunnel inelastically exciting the optical (in most cases plasmonic) modes of the structure. With an appropriate design of the optical modes, upon their coupling to free-space radiation, the intensity of the emitted light can be so high, that the emission is visible to the naked eye \cite{wang2018reactive}. Through engineering of the spatial and spectral profile of the MIM plasmonic modes, the directional and polarisation properties of the emitted light can be efficiently controlled \cite{kirtley1981light,lee2025plasmonic}. However, the electric generation of chiral light in the tunnel junctions so far remains elusive.   

Here, at a single particle level we demonstrate electrically-driven nanoscale chiral light emission from tunnel junctions incorporating chiral plasmonic nanohelicoids. Supporting resonant chiral dipolar modes, such nanostructures generate high chiral LDOS in the tunnelling gap, imprinting chirality into the tunnelling-driven light emission. The experimentally measured emission dissymmetry factor between the right-hand and left-hand polarisation of the emitted light reaches values as high as 0.8. Furthermore, obeying the angular momentum conservation, the emitted light also carries an orbital angular momentum, of an equal magnitude and opposite sign, confirmed by numerical modelling. Such vortex beams are intrinsically difficult to generate or focus at subwavelength scales \cite{Litchinitser2025vortex}, while they are important, e.g., for excitation of dark molecular transitions, optical trapping and spintronic applications.  
The demonstrated electrically-driven chiral light emission from nanoscale tunnel junctions opens up new avenues for exploration of fundamental chiral light-matter interactions in the quantum limit and offers  a promising technological platform for the realisation of ultra-compact chiral light sources for chiral nanophotonics, integrated quantum photonic circuits and AR/VR displays.

\section{Results}
\subsection{Design of electrically-driven chiral tunnel junctions}

\begin{figure}[!ht]
    \begin{center}
        \includegraphics[width=14cm]{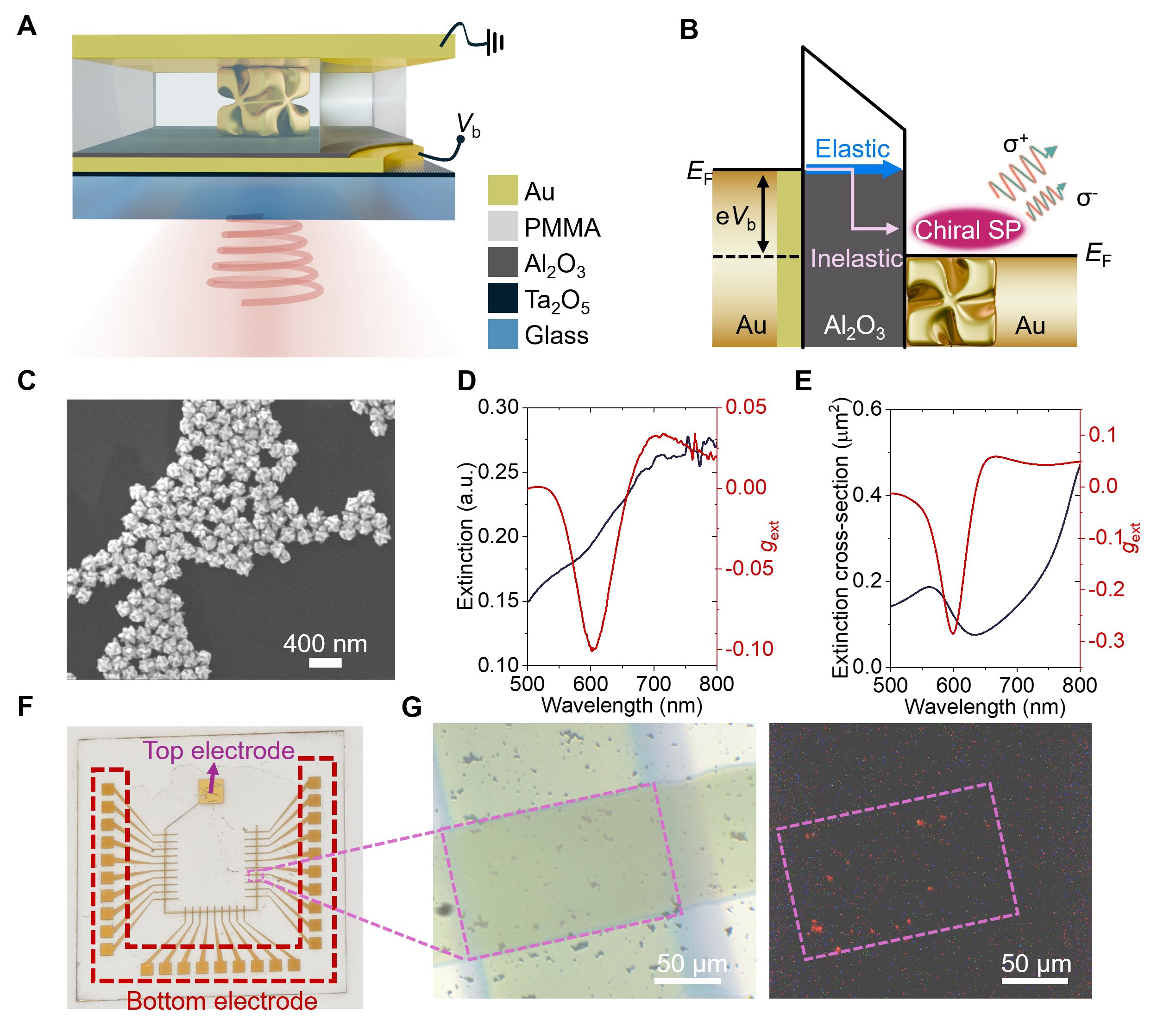}
        \caption{\textbf{Chiral light emission from plasmonic chiral tunnel junctions.} \textbf{a},~Schematics of a chiral tunnelling device based on a plasmonic helicoid. \textbf{b},~Energy-level diagram and schematics of the inelastic electron tunnelling pathway, leading to the excitation of chiral plasmonic modes coupled to chiral light emission. \textbf{c},~SEM image of the fabricated helicoid nanoparticles. \textbf{d},~\textbf{e},~Experimentally measured (\textbf{d}) and numerically simulated (\textbf{e}) extinction (black lines) and extinction \textit{g}-factors ($g_\text{ext}$) (red lines) spectra of the helicoid nanoparticles in a water solution. \textbf{f},~ Optical photograph of an array of tunnelling devices containing ensembles of helicoid nanoparticles used in the experiments. \textbf{g},~ Zoomed-in optical image of an individual tunnelling device showing (left) the set of single tunnel junctions within it and (right) the emission under an applied bias of 2.2~V.}\label{fig1}
    \end{center}
\end{figure}

The plasmonically-assisted chiral tunnel junctions were designed by integrating electrically-contacted gold nanohelicoids separated from a gold film electrode by an alumina dielectric layer, forming the tunnelling gap  (see schematics in Fig.~\ref{fig1}a,b and Methods for the details of fabrication). The gold helicoid nanoparticles (Fig.~\ref{fig1}c) serve as chiral plasmonic resonators due to their pronounced chiroptical response characterised by the extinction g-factor ($g_\text{ext}=2(I_\text{ext}^\text{LCP}-I_\text{ext}^\text{RCP})/(I_\text{ext}^\text{LCP}+I_\text{ext}^\text{RCP})$), observed experimentally (Fig.~\ref{fig1}d) and confirmed numerically (Fig.~\ref{fig1}e).

During fabrication, the nanohelicoids were dispersed on the 30-nm-thick gold film covered with the 3-nm-thick \ch{Al2O_3} barrier layer and encapsulated into a PMMA layer. The PMMA was subsequently plasma-milled to expose the helicoid top facets (see Supplementary Text
), after which a 50-nm thick Au layer was deposited to serve as the top electrode. The polymer matrix provides a suitable dielectric environment to optimise the helicoid chiral response, which depends on the refractive index of surroundings \cite{xie2025unidirectional}, and, crucially, acts as an insulating spacer to prevent direct electron tunnelling between the top and the bottom gold electrodes. This design ensures that the detected light emission originates exclusively from the chiral tunnel junctions produced by the nanohelicoids (Fig.~\ref{fig1}g). Application of a bias voltage ($V_\text{b}$) between the helicoid and the bottom gold film initiates an electron tunnelling process (Fig.~\ref{fig1}b). The majority of electrons tunnel elastically conserving their energy, but a small fraction of them undergoes inelastic tunnelling, which leads to the excitation of chiral surface plasmon modes supported by the nanohelicoid-film structure. Upon coupling to the free-space radiation, these surface plasmon modes produce the emission of chiral light possessing both SAM and OAM. The spectrum of the chiral emission is defined by the product of the spectral profile of the voltage-dependent tunnelling sources  $I_{\text{tc}}(\omega,V_\text{b})\propto(1-\hbar\omega/eV_\text{b})$ (determined by a Fourier transform of the tunnelling current fluctuations \cite{Rendell1981}), the spectrum of the supported chiral plasmonic modes (most importantly their local density of optical states in the tunnelling gap where the sources are located), and the coupling efficiencies of the modes to the free-space radiation \cite{wang2018reactive,parzefall2015antenna}.

\subsection{Chiral electro-optical properties}

\begin{figure}[!ht]
    \begin{center}
        \includegraphics[width=14cm]{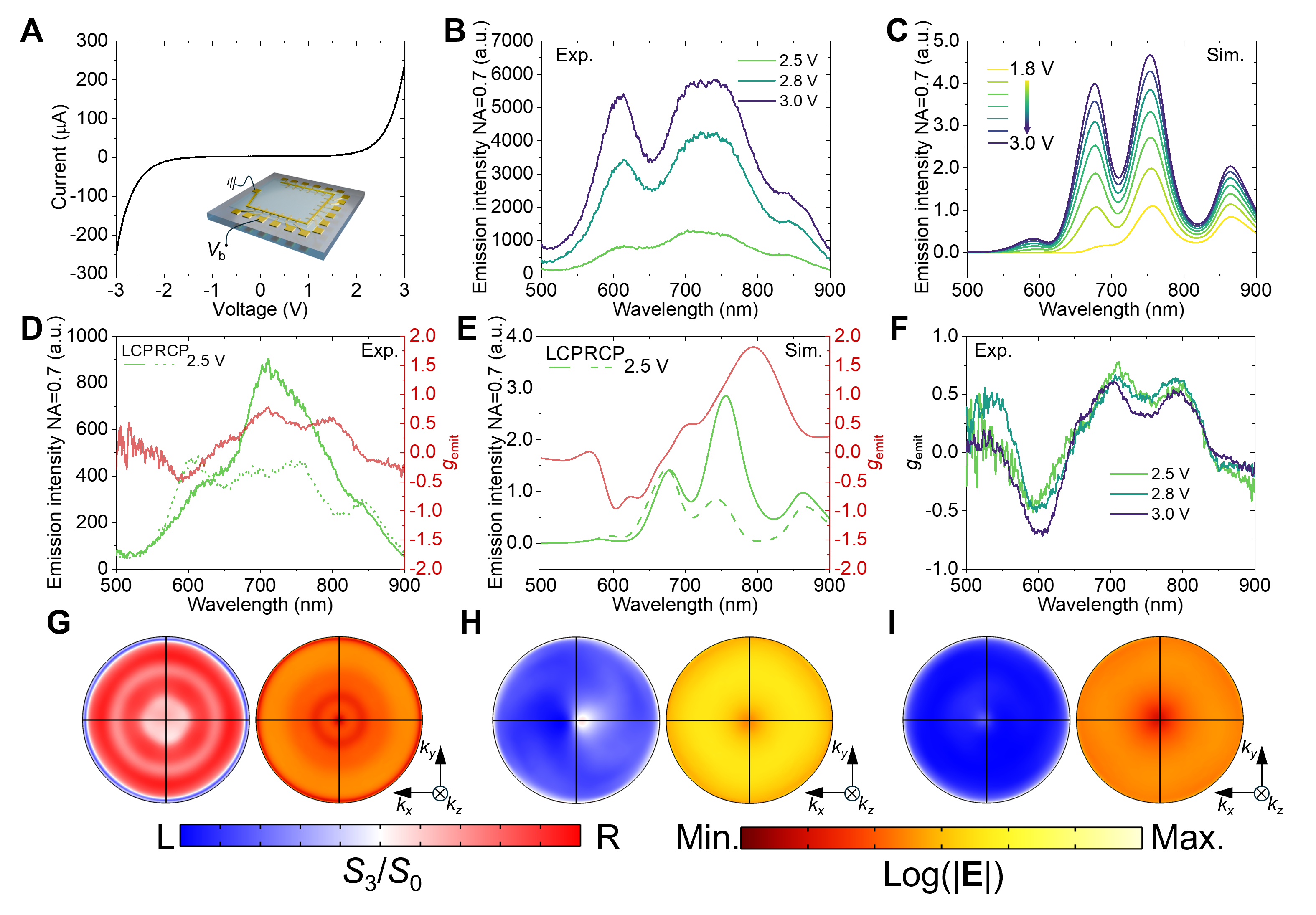}
        \caption{\textbf{Tunnelling-driven resonant excitation of chiral plasmonic modes.} \textbf{a},~Experimentally measured I-V curve for one of the tunnelling devices shown in Fig.~\ref{fig1}f.
        \textbf{b},~\textbf{c},~ Experimentally measured (\textbf{b}) and numerically simulated (\textbf{c})
        emission spectra from a single nanohelicoid chiral tunnel junction under a 2.5~V forward bias, collected inside a $\text{NA}=0.70$ solid angle form the substrate side. \textbf{d},~\textbf{e},~ The corresponding experimentally measured (\textbf{d}) and numerically simulated (\textbf{e}) emission spectra of LCP (solid green lines) and RCP (dashed green lines) components of the emitted light, together with the related spectral dependence of the emission \textit{g}-factor  ($g_\text{emit}$) (red lines) in the case of a 2.5~V forward bias. Following the definition adopted in CD spectrometry, the light handedness is defined from the point of view of a receiver. In numerical simulations the \textbf{f},~Experimentally measured $g_\text{emit}$ spectra under various applied biases. \textbf{g--i},~ Numerically simulated maps of light handedness (left panels) and field intensity (right panels) of the emitted light for the wavelengths of 600~nm (\textbf{g}), 760~nm (\textbf{h}) and 800~nm (\textbf{i}). Very small intensity for the emission close to the normal direction leads to relatively large numerical uncertainty in the calculation of the polarisation state, which leads to slight asymmetry of the polarisation maps in this region.}\label{fig2}
    \end{center}
\end{figure}

The measured current–voltage (I–V) curves of the fabricated devices serves as a direct indicator of the initiated tunnelling process (Fig.~\ref{fig2}a). Unlike a linear I-V dependence of an Ohmic contact or an asymmetric dependence under forward and reverse bias of its Schottky counterpart,  
the tunnelling contact exhibits a nonlinear but symmetric I–V characteristic when the same metal forms both electrodes. This distinct signature provides an effective diagnostic tool to verify the integrity of the tunnel junction during measurements. 

Spectral dependences of an emission collected from a single chiral junction under various applied biases are in good agreement with numerical predictions (Fig.~\ref{fig2}b,c). 
Three pronounced emission peaks experimentally observed at the 620~nm, 740~nm and 860~nm wavelengths correspond to resonantly excited plasmonic modes of the nanostructure. The slight difference in peak positions in the experimental and simulation results may be caused by the thickness variations of the bottom gold layer: as was numerically observed, the increased thickness of the bottom gold layer results in a pronounced blue shift of the peaks accompanied by attenuation of their magnitude (see Supplementary Text). 
As the bias increases, the emission spectrum not only intensifies but also reshapes: the enhancement is more pronounced at shorter wavelengths, resulting in an overall blue shift of the spectral profile in both experiment and simulations (Fig.~\ref{fig2}b,c). The variation of the spectra is related to the change in the spectral profile of the source tunnelling current $I_{\text{tc}}(\omega,V_\text{b})$ and through this to the change of the excitation probability of different plasmonic modes.

The emitted light was then separated into its left- and right-handed circularly polarised (LCP and RCP) components using a quarter-wave plate (QWP) and a calcite beam displacer (CBD). 
Fig.~\ref{fig2}d presents the resulting spectral dependences of LCP and RCP emission intensities, as well as the corresponding emission polarisation asymmetry factor, defined as $g_\text{emit}=2(I_\text{emit}^\text{LCP}-I_\text{emit}^\text{RCP})/(I_\text{emit}^\text{LCP}+I_\text{emit}^\text{RCP})$. Both experimental and numerically simulated $g_\text{emit}$ spectra exhibit the same trend across the measured spectral range (Fig.~\ref{fig2}d,e). The experimental $g_\text{emit}$ reaches a maximum value of $\simeq$0.8, demonstrating excellent chiral emission performance. Such large $g_\text{emit}$ value primarily arises from the suppression of the RCP emission peak around 760~nm, as predicted in the numerical simulations (Fig.~\ref{fig2}e). While the intrinsic $g_\text{emit}$ spectrum should be independent of the applied bias voltage (although the spectral dependence of the total emitted power changes, the ratio between the RCP and LCP components at any given wavelength remains the same), experiment replicated under varying bias conditions show minor spectral fluctuations (Fig.~\ref{fig2}f). These slight variations may originate from fluctuations in the handedness distribution across different emission directions (Fig.~\ref{fig2}g--i).
The simulated emission patterns at the positions of $g_\text{emit}$ extrema (at 600 and 800~nm wavelengths) and the maximum emission intensity (760~nm wavelength) have a well defined cylindrical symmetry with distinct circularly polarised polarisation states (Fig.~\ref{fig2}g--i).

\subsection{Origin of chirality of emitted light}

\begin{figure}[!ht]
    \begin{center}
        \includegraphics[width=14cm]{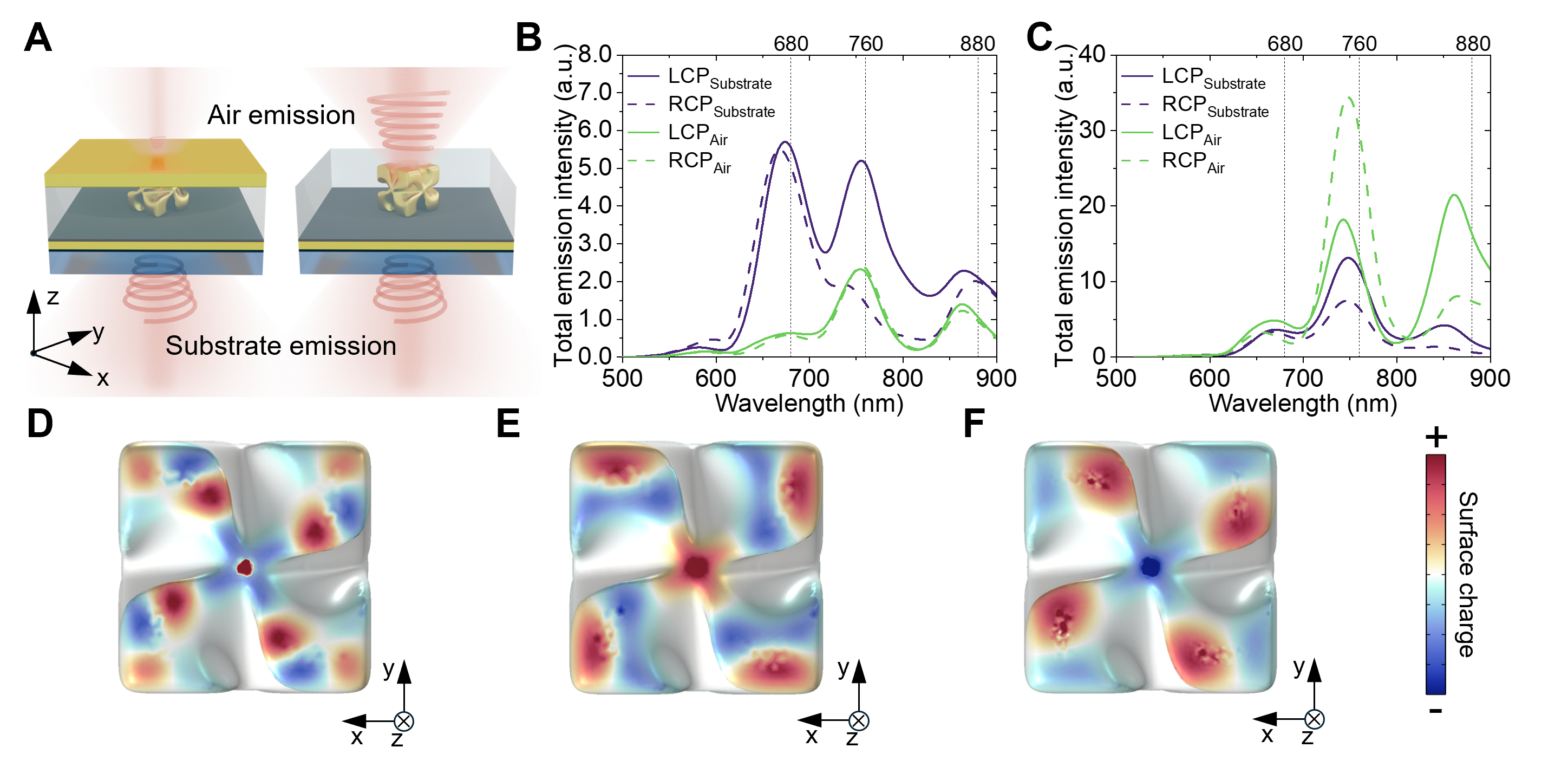}
        \caption{\textbf{Numerically simulated near-field origin of the plasmonically-assisted chiral emission.} \textbf{a},~Schematics of emission pathways into air and substrate with and without a top gold electrode. \textbf{b},~\textbf{c},~The simulated emission spectra for LCP (solid lines) and RCP (dashed lines) components emitted into the air hemisphere (green lines) and the substrate hemisphere (purple lines) with (\textbf{b}) and without (\textbf{c}) the top gold electrode. \textbf{d--f}~The surface charge distributions on the tunnelling facet of the helicoid nanoparticle for the wavelengths corresponding to the emission peaks of the experimentally studied nanostructures (with the top electrode) 680~nm (\textbf{d}), 760~nm (\textbf{e}) and 880~nm (\textbf{f}), shown in Fig.~\ref{fig2}c.}\label{fig3}
    \end{center}
\end{figure}

The emitted light exhibits dissimilar chiral behaviour at the three emission peaks. Notably, only the middle peak (around 760~nm) shows a prominent chiral response, with the polarisation asymmetry factor ranging  $g_\text{emit}^\text{760 nm}>g_\text{emit}^\text{880 nm}>g_\text{emit}^\text{680 nm}$. To understand these differences, the emission into both air and substrate sides of the device was numerically simulated (Fig.~\ref{fig3}a). When an optically thick 50-nm top gold electrode is present, it acts as a mirror and strongly suppresses the emission into the air. Particularly, compared to the emission into the substrate, the resulting emission in the air has much weaker intensity and a reduced handedness contrast, but at the same time it retains the three spectral peaks structure (Fig.~\ref{fig3}b).

In contrast, upon removal of the top gold electrode in the simulated model, emission into both air and substrate become more pronounced, displaying a strong chiral response (Fig.~\ref{fig3}c), which makes the different behaviour among the three peaks clearer: 1)~at the $\sim$680~nm peak, emission signals both into the air and the substrate show weak chirality, 2)~at the $\sim$760~nm peak, the the emission signals in the air and in the substrate possess opposite dominating handedness, 3)~at the $\sim$880~nm peak, LCP light dominates emission into both directions, yielding the strongest combined chiral response. These simulation results clarify the experimental observations. In particular, for the device with the top gold layer, at a 760~nm wavelength, the originally RCP-dominated emission into the air (in the model without the top electrode, see Fig.~\ref{fig3}C) is reflected back into the substrate direction with an expected change of handedness. This constructively enhances the LCP emission into the substrate at this wavelength, leading to the observed maximum in $g_\text{emit}$. Conversely, at the 860~nm wavelength, for which the original emission in both directions have the same handedness (LCP, see Fig.~\ref{fig3}C), this flip of handedness causes a decrease of the purity of circular polarisation state emitted into the substrate, thereby diminishing the chiral response.

The observed three emission peaks originate from the plasmonic modes of the nanostructure, while the numerically observed independence of the spectral positions of these modes on the presence of the top gold electrode suggests that by nature they are predominantly MIM gap modes localised in the tunnelling junction for which the bottom facet of the helicoid creates an in-plane resonator. The excitation and predominance of this modes is logical given that the excitation sources produced by the tunnelling current are located exactly in the gap.
The distinct chiral responses at the spectral position of each peak are linked to different resonant plasmonic modes (cf. Fig.~\ref{fig2}b and d, and Fig.~\ref{fig2}c and e). Analysis of the surface charge distributions on the helicoid bottom facet at the wavelength of each peak (Fig.~\ref{fig3}d--f) reveals that the gap mode couples into distinct curved regions of the helicoid facet. At the same time, for all three modes, the 4-fold rotational symmetry of the helicoid is translated into the corresponding symmetry of the plasmonic fields. The three resonant modes are expectedly localised at the helicoid bottom lobes and in the region of their connection at the centre, in the areas of the smallest gaps between the helicoid and the bottom layer. At the same time, the order of these modes is clearly different, with four field extrema per lobe at 680~nm, three at 760~nm, and two at 880~nm, logically decreasing as the wavelength increases.
The difference in the spatial distribution of the plasmonic modes is converted into their different chiral response. The role of the helicoid chirality for the phenomenon of chiral emission was confirmed by modelling tunnelling-assisted emission from a non-chiral cube of an identical size. A similar three-peak emission spectrum was observed, but with a zero chiral response within the accuracy of the numerical simulations (see Supplementary Text for details). This contrast underscores that the chiral morphology is central for generating the pronounced chiral emission.

\subsection{Tunnelling-driven generation of orbital angular momentum}
\begin{figure}[!ht]
    \begin{center}
        \includegraphics[width=16cm]{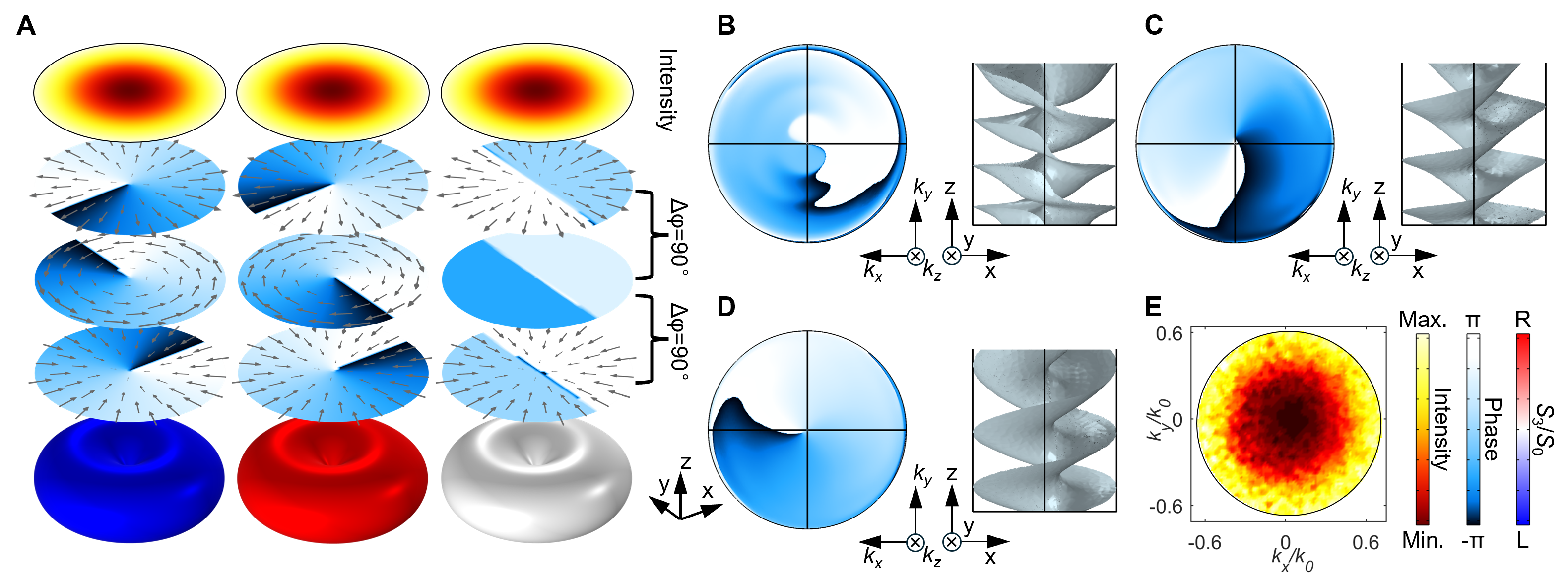}
        \caption{\textbf{Vorticity of light generated from chiral tunnel junction.} \textbf{a},~Comparison of OAM generation from chiral dipoles (OAM generated) and a linear electric dipole (an equal in-phase sum of LCP and RCP chiral dipoles, no OAM generated). The fieldmaps represent, from top to bottom: (1) intensity, (2--4) phase evolution at discrete time intervals with arrows showing the direction of electric field vectors, and (5) radiation patterns with the colour indicating the polarisation state of light. \textbf{b}--\textbf{d},~Numerically simulated distributions of cross-sectional phase distribution (left) and 3D constant-phase wavefronts for light emitted into the substrate, evaluated at the wavelengths of 600~nm (\textbf{b}), 760~nm (\textbf{c}) and 800~nm (\textbf{d}).  \textbf{e},~Fourier-plane intensity distribution for light emitted from of the electrically-driven chiral tunnel junction, measured with a $750\pm40$~nm bandpass filter (corresponding to averaged $g_\text{emit}=0.48$) under a 3~V forward bias.}\label{fig4}
    \end{center}
\end{figure}

Extending beyond the circular polarised handedness of the field (related to the spin angular momentum, SAM), the chirality of a beam also encompasses the handedness of its wavefront rotation around its propagation axis, associated with the orbital angular momentum, OAM. In the case of the tunnelling-driven excitation of helicoid nanostructures, most of the emission comes from dipolar plasmonic modes, particularly corresponding to electric ($\textbf{p}$) and magnetic ($\textbf{m}$) dipoles directed along the $z$-axis, which can be seen from multipolar decomposition of the nanohelicoid excited currents presented in Fig.~S9. Combined together with a certain magnitude ratio $p = m/c$ and a $\pi/2$ ($-\pi/2$) phase shift ($\text{exp}(-i\omega t)$ convention is used throughout the paper), co-directed electric and magnetic dipoles produce a chiral dipole, emitting light in a usual dipolar pattern, but having RCP- (LCP-) polarisation \cite{eismann2018exciting}. If the magnitude or phase relations are not strictly satisfied, the emission will be defined by a sum of RCP- and LCP- chiral dipoles (Fig.~S9), and have both circular components in its polarisation, so the emission will be predominantly, but not purely RCP or LCP. Such nearly-pure states of chiral dipoles with the handedness depending on the wavelength and defined by the excited resonant plasmonic mode was realised using tunnelling-driven excitation in the helicoid nanostructures (Fig.~\ref{fig2}g--i). 
At the same time, the emission from chiral dipoles in addition to a spin angular momentum (observed as its polarisation state) must also carry an orbital angular momentum, due to the angular momentum conservation. Specifically, ideal RCP (LCP) chiral dipoles emit light with an orbital angular momentum $l = 1$ ($l = -1$) along their axis, which can be clearly seen in the maps of the phase variation along the dipolar axis (Fig.~\ref{fig4}a). A combination of RCP and LCP chiral dipoles emits a mixture of $l = 1$ and $l = -1$ OAM states (Fig.~\ref{fig4}a and Fig.~S3).
In the case of the observed tunnelling-driven emission, the nearly-pure chiral dipole emit light with predominant $l = 1$ or $l = -1$ OAM. This conclusion is fully confirmed by the numerical simulation results presented in Fig.~\ref{fig4}b--d, where the $0-\pm2\pi$ variation of the phase around the beam axis for the three wavelengths corresponding to the extrema of the nanostructure chiral response (600 and 800~nm, Fig.~\ref{fig2}e) and the emission intensity (760~nm) (cf. a phase map for a highly mixed chiral dipole state in Fig.~S10, which correspond to weak chiral emission of intermediate dipolar states in Fig.~S3, for which 2$\pi$ phase variation is not achieved). Comparing the direction of the rotation of the phase (related to OAM) and the handedness of the emission (related to SAM), one can conclude that there is momentum conservation in the emission from an initially achiral tunnelling event (described by a $z$-oriented electric dipole) to the light having equal and opposite in sign OAM and SAM. The dipolar nature of the emission was observed numerically (Fig.~\ref{fig2}g--i) and confirmed experimentally in spacial distribution of the emitted intensity from a tunnelling-driven single nanohelicoid structure measured using Fourier microscopy (Fig.~\ref{fig4}e). This confirms that chiral tunnel junctions provide a novel pathway toward achieving electrically-driven vortex beam emission at the nanoscale. 

\clearpage

\section{Conclusions}

We have developed a novel platform for electrically-driven generation of chiral light at the nanoscale via electron tunnelling by integrating chiral plasmonic nanoparticles with metal-insulator-metal tunnel junctions. Leveraging the excitation of chiral plasmonic gap modes, broadband chiral light emission was demonstrated in the visible spectral range at the single-particle level. The resulting circular polarisation asymmetry reached values up to 0.8, representing a significant advancement in polarisation control of light emission at the nanoscale. Furthermore, such emission, driven by the excitation of chiral dipolar modes supported by the nanostructure, simultaneously results in generation of an orbital angular momentum in the emitted light, equal and opposite in sign to the spin angular momentum, related to the handedness.
These findings bridge the gap between quantum tunnelling processes and chiral nanophotonics, providing a robust strategy for active polarisation management without the need for bulky external optics. The developed platform holds substantial promise for the development of next-generation optoelectronic devices, with potential applications ranging from high-resolution displays and AR/VR technologies to quantum information processing and enantioselective photochemistry.

\section{References}
\bibliographystyle{naturemag}
\bibliography{bib}
\section*{Acknowledgements}
The authors are grateful to Prof.~Ki Tae Nam for providing a CAD geometry of the nanohelicoid (432 helicoid III) used in the simulations; Dr. Anastasiia Zaleska for the help with sample preparation; Dr.Tam Bui and Optical Spectroscopy Facility, King's College London for CD spectra measurements. 

\section*{Research funding}

This work was supported by the UK EPSRC projects EP/W017075/1 and UKRI3056. 

\section*{Author contributions}
Y.X., A.V.K and A.V.Z. developed the idea, Y.X. and A.V.K. performed the numerical simulations, Y.X prepared the samples and performed the measurements, all authors contributed to the writing of the manuscript. 

\end{document}